\documentclass[3p,times,twocolumn]{elsarticle}

\usepackage{ecrc}

\volume{00}

\firstpage{1}

\journalname{Nuclear Physics B Proceedings Supplement}

\runauth{E. Goudzovski}

\jid{nuphbp}

\jnltitlelogo{Nuclear Physics B Proceedings Supplement}

\usepackage{amssymb}

\usepackage[figuresright]{rotating}

\begin{document}

\begin{frontmatter}



\dochead{}

\title{Kaon experiments at CERN: NA48 and NA62}


\author{Evgueni Goudzovski}

\address{School of Physics and Astronomy, University of Birmingham,
B15 2TT, United Kingdom}

\begin{abstract}
Searches for violation of lepton flavour universality and lepton
number conservation in kaon decays by the NA62 and NA48/2
experiments at CERN, status and future plans of the CERN kaon
programme are presented. A precision measurement of the
helicity-suppressed ratio $R_K$ of the $K^\pm\to e^\pm\nu$ and
$K^\pm\to\mu^\pm\nu$ decay rates has been performed using the full
data set collected by the NA62 experiment in 2007--2008. The result
is $R_K=(2.488\pm0.010)\times10^{-5}$, in agreement with the
Standard Model expectation. An improved upper limit on the rate of
the lepton number violating $K^\pm\to\pi^\mp\mu^\pm\mu^\pm$ decay
from the NA48/2 experiment (2003--2004 data set) is presented.
Finally, the NA62 project aiming at a measurement of the branching
ratio of the ultra-rare decay $K^+\to\pi^+\nu\bar\nu$ at 10\%
precision is discussed.

\end{abstract}


\end{frontmatter}


\section{Introduction}
\label{intro}

Decays of pseudoscalar mesons to light leptons
($P^\pm\to\ell^\pm\nu$, denoted $P_{\ell 2}$ in the following) are
suppressed in the Standard Model (SM) by helicity considerations.
Specific ratios of leptonic decay rates can be computed very
precisely: in particular, the SM prediction for the ratio
$R_K=\Gamma(K_{e2})/\Gamma(K_{\mu 2})$ is~\cite{ci07}
\begin{eqnarray*}
\label{Rdef} R_K^\mathrm{SM} &=& \left(\frac{m_e}{m_\mu}\right)^2
\left(\frac{m_K^2-m_e^2}{m_K^2-m_\mu^2}\right)^2 (1 + \delta
R_{\mathrm{QED}})=\\
&=&(2.477 \pm 0.001)\times 10^{-5},
\end{eqnarray*}
where $\delta R_{\mathrm{QED}}=(-3.79\pm0.04)\%$ is an
electromagnetic correction. Within certain two Higgs doublet models
(2HDM type II), $R_K$ is sensitive to lepton flavour universality
violating effects via the charged Higgs boson ($H^\pm$)
exchange~\cite{ma08}, the dominant contribution being
\begin{displaymath}
\frac{R_K^\mathrm{}}{R_K^\mathrm{SM}}\simeq
1+\left(\frac{m_K}{m_H}\right)^4 \left(\frac{m_\tau}{m_e}\right)^2
|\Delta _R^{31}|^2\tan^6\beta, \label{rk_lfv}
\end{displaymath}
where $\tan\beta$ is the ratio of the two Higgs vacuum expectation
values, and $|\Delta_{R}^{31}|$ is the mixing parameter between the
superpartners of the right-handed leptons, which can reach $\sim
10^{-3}$. This can enhance $R_K$ by ${\cal O}(1\%)$ without
contradicting any known experimental constraints. A precise $R_K$
measurement based on the full data sample collected during the $R_K$
phase of the CERN NA62 experiment in 2007--2008 ($\sim 10$ times the
world $K_{e2}$ sample) is presented, superseding an earlier
result~\cite{la11} based on a partial data set.

The decay $K^\pm\to\pi^\mp\mu^\pm\mu^\pm$ violating lepton number
conservation by two units can proceed via a neutrino exchange if the
neutrino is a Majorana particle~\cite{zu00}; it has also been
studied in the context of supersymmetric models with $R$-parity
violation~\cite{li00}. The most stringent limits on this and similar
lepton flavour and number violating processes come from the BNL E865
experiment~\cite{ap00}. The NA48/2 experiment at CERN collected the
world's largest $K^\pm\to\pi\mu\mu$ sample in 2003--2004, which
allowed a significant improvement of the BNL upper limit on ${\rm
BR}(K^\pm\to\pi^\mp\mu^\pm\mu^\pm)$.

Among the flavour changing neutral current $K$ and $B$ decays, the
ultra rare decays $K\to\pi\nu\bar\nu$ play a key role in the search
for new physics through the underlying mechanisms of flavour mixing.
These decays are strongly suppressed in the SM (highest CKM
suppression), and are dominated by top-quark loop contributions. The
SM branching ratios have been computed to an exceptionally high
precision with respect to other loop-induced meson decays: ${\rm
BR}(K^+\to\pi^+\nu\bar\nu)=8.22(75)\times 10^{-11}$ and ${\rm
BR}(K_L\to\pi^0\nu\bar\nu)=2.57(37)\times 10^{-11}$; the
uncertainties are dominated by parametric ones, and the irreducible
theoretical uncertainties are at a $\sim 1\%$ level~\cite{br11}. The
extreme theoretical cleanness of these decays remains also in new
physics scenarios like Minimal Flavour Violation (MFV) or non-MFV
models, and even modest deviations of BRs from their SM values can
be considered signals of new physics.

The $K^+\to\pi^+\nu\bar\nu$ decay has been observed by the E787/E949
experiments at the BNL, and the measured branching ratio is
$\left(1.73^{+1.15}_{-1.05}\right)\times 10^{-10}$~\cite{ar09}. The
achieved precision is inferior to that of the SM expectation. The
main goal of the NA62 experiment at CERN is the measurement of the
$K^+\to\pi^+\nu\bar\nu$ decay rate at the 10\% precision level,
which would constitute a significant test of the SM.

\section{Search for lepton flavour universality violation}

\begin{figure*}[tb]
\begin{center}
\resizebox{0.5\textwidth}{!}{\includegraphics{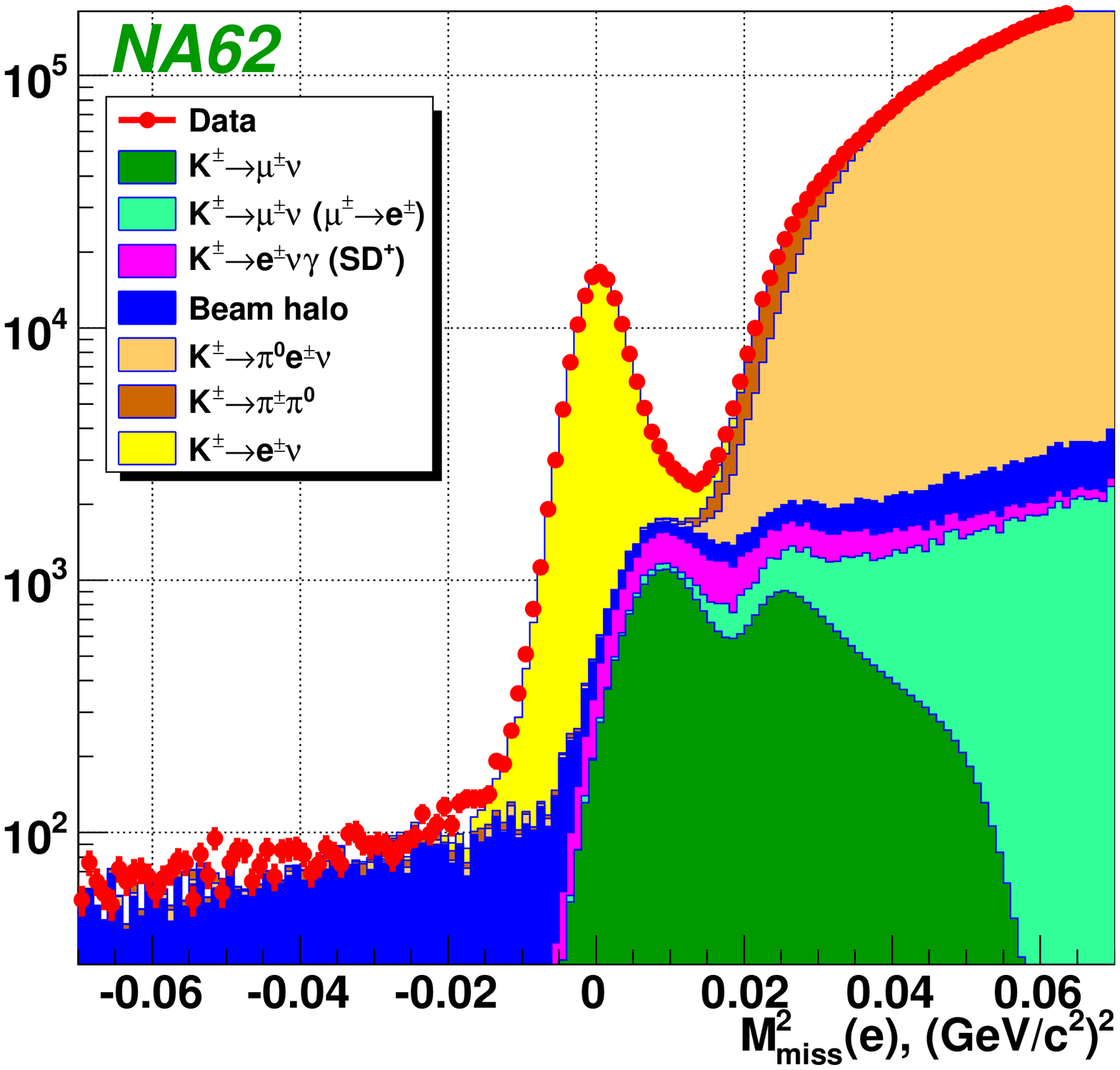}}%
\resizebox{0.5\textwidth}{!}{\includegraphics{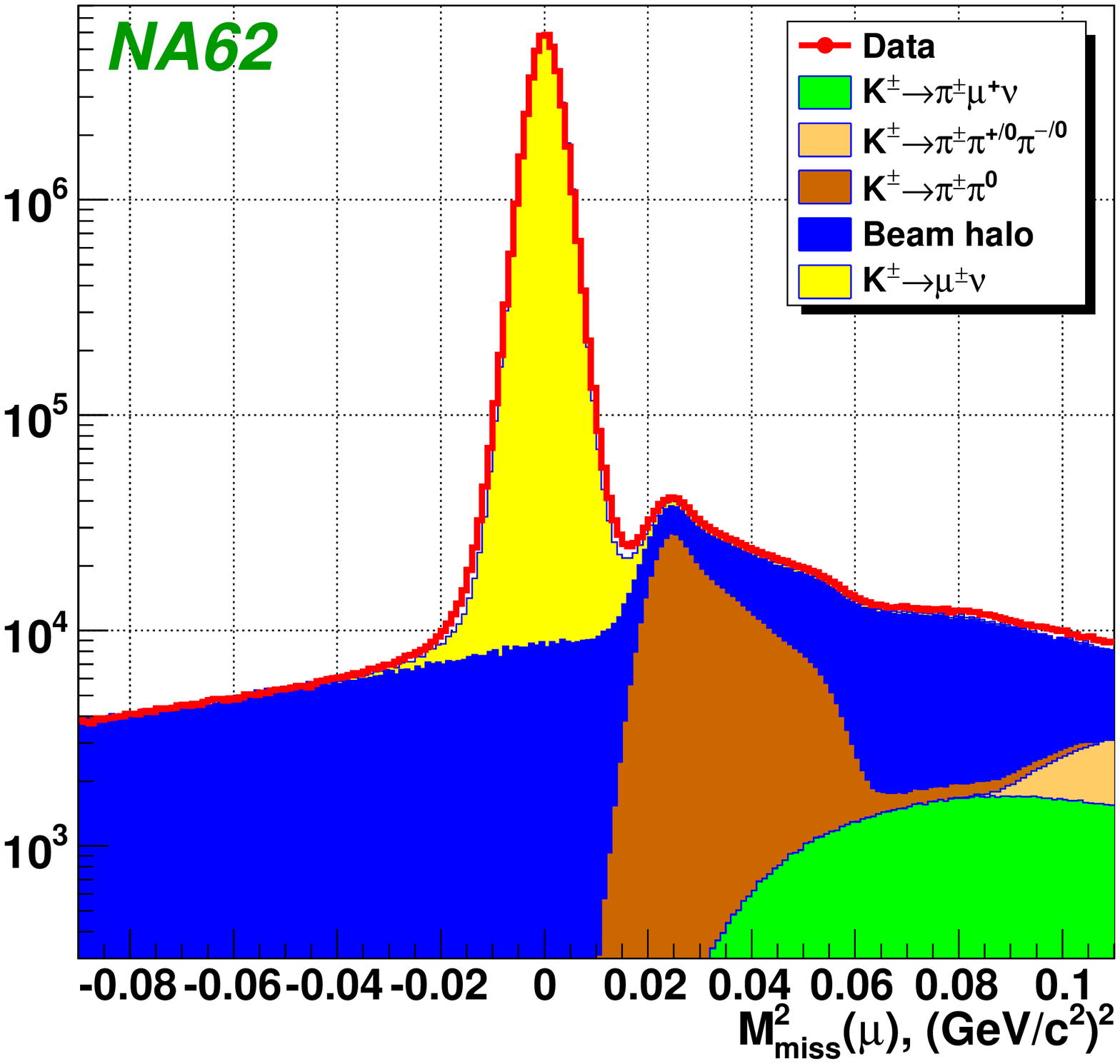}}%
\end{center}
\vspace{-6mm} \caption{Distributions of the reconstructed squared
missing mass $M_{\mathrm{miss}}^2(e)$ (left) and
$M_{\mathrm{miss}}^2(\mu)$ (right) of the $K_{e2}$ ($K_{\mu 2}$)
candidates compared with the sums of normalised estimated signal and
background components. The double peak structure in the spectrum of
the $K_{\mu 2}$ background to $K_{e2}$ decay is due to the momentum
dependent particle identification criterion.} \label{fig:mm2}
\end{figure*}

The beam line and setup of the NA48/2 experiment~\cite{fa07} have
been used during the NA62 $R_K$ phase. The beam line was originally
designed to deliver simultaneous narrow momentum band $K^+$ and
$K^-$ beams derived from the primary 400 GeV/$c$ protons extracted
from the CERN SPS. In 2007, mostly single beam configurations (with
74 GeV/$c$ central momentum and 1.4~GeV/$c$ rms spread) were used.
The momenta of individual beam particles were not measured. The beam
kaons decayed in a fiducial decay volume contained in a 114~m long
cylindrical vacuum tank.

The momenta of charged decay products are measured in a magnetic
spectrometer, housed in a tank filled with helium placed after the
decay volume. The spectrometer comprises four drift chambers (DCHs),
two upstream and two downstream of a dipole magnet which gives a
horizontal transverse momentum kick of $265~\mathrm{MeV}/c$ to
singly-charged particles. Each DCH is composed of eight planes of
sense wires. The measured spectrometer momentum resolution is
$\Delta p/p = 0.48\% \oplus 0.009\% p$, where the momentum $p$ is
expressed in GeV/$c$. A plastic scintillator hodoscope (HOD)
producing fast trigger signals and providing precise time
measurements of charged particles is placed after the spectrometer.
A 127~cm thick liquid krypton (LKr) electromagnetic calorimeter
located further downstream is used for lepton identification and as
a photon veto detector. Its 13248 readout cells have a transverse
size of approximately 2$\times$2 cm$^2$ each, without longitudinal
segmentation.

The analysis strategy is based on counting the numbers of
reconstructed $K_{e2}$ and $K_{\mu 2}$ candidates collected
concurrently. Therefore the analysis does not rely on the absolute
beam flux measurement, and several systematic effects cancel at
first order. The study is performed independently for 40 data
samples (10 bins of reconstructed lepton momentum and 4 samples with
different data taking conditions) by computing the ratio $R_K$ as
\begin{eqnarray*}
R_K &=& \frac{1}{D}\cdot \frac{N(K_{e2})-N_{\rm
B}(K_{e2})}{N(K_{\mu2}) - N_{\rm B}(K_{\mu2})}\cdot
\frac{A(K_{\mu2})}{A(K_{e2})} \cdot\\
&&\frac{f_\mu\times\epsilon(K_{\mu2})}
{f_e\times\epsilon(K_{e2})}\cdot\frac{1}{f_\mathrm{LKr}},
\label{eq:rkcomp}
\end{eqnarray*}
where $N(K_{\ell 2})$ are the numbers of selected $K_{\ell 2}$
candidates $(\ell=e,\mu)$, $N_{\rm B}(K_{\ell 2})$ are the numbers
of background events, $A(K_{\mu 2})/A(K_{e2})$ is the geometric
acceptance correction, $f_\ell$ are the efficiencies of $e$/$\mu$
identification, $\epsilon(K_{\ell 2})$ are the trigger efficiencies,
$f_\mathrm{LKr}$ is the global efficiency of the LKr calorimeter
readout (which provides the information used for electron
identification), and $D$ is the downscaling factor of the $K_{\mu2}$
trigger. A Monte Carlo simulation is used to evaluate the acceptance
correction and the geometric part of the acceptances for background
processes entering the computation of $N_B(K_{\ell 2})$. Particle
identification, trigger and readout efficiencies and the beam halo
background are measured directly from control data samples.

Two selection criteria are used to distinguish $K_{e2}$ and
$K_{\mu2}$ decays. Kinematic identification is based on the
reconstructed squared missing mass assuming the track to be a
electron or a muon: $M_{\mathrm{miss}}^2(\ell) = (P_K - P_\ell)^2$,
where $P_K$ and $P_\ell$ ($\ell = e,\mu$) are the kaon and lepton
4-momenta (Fig.~\ref{fig:mm2}). A selection condition
$M_1^2<M_{\mathrm{miss}}^2(\ell)<M_2^2$ is applied; $M_{1,2}^2$ vary
across the lepton momentum bins depending on resolution. Lepton type
identification is based on the ratio $E/p$ of energy deposit in the
LKr calorimeter to track momentum measured by the spectrometer.
Particles with $(E/p)_{\rm min}<E/p<1.1$ ($E/p<0.85$) are identified
as electrons (muons). Here $(E/p)_{\rm min}$ is 0.90 or 0.95,
depending on momentum.

The largest background to the $K_{e2}$ decay is the $K_{\mu2}$ decay
with a mis-identified muon ($E/p>0.95$) via the `catastrophic'
bremsstrahlung process in the LKr. To reduce the corresponding
uncertainty, the muon mis-identification probability $P_{\mu e}$ has
been measured as a function of momentum using dedicated data
samples.

\begin{figure*}[tb]
\begin{center}
\vspace{-2mm}
\resizebox{0.6\textwidth}{!}{\includegraphics{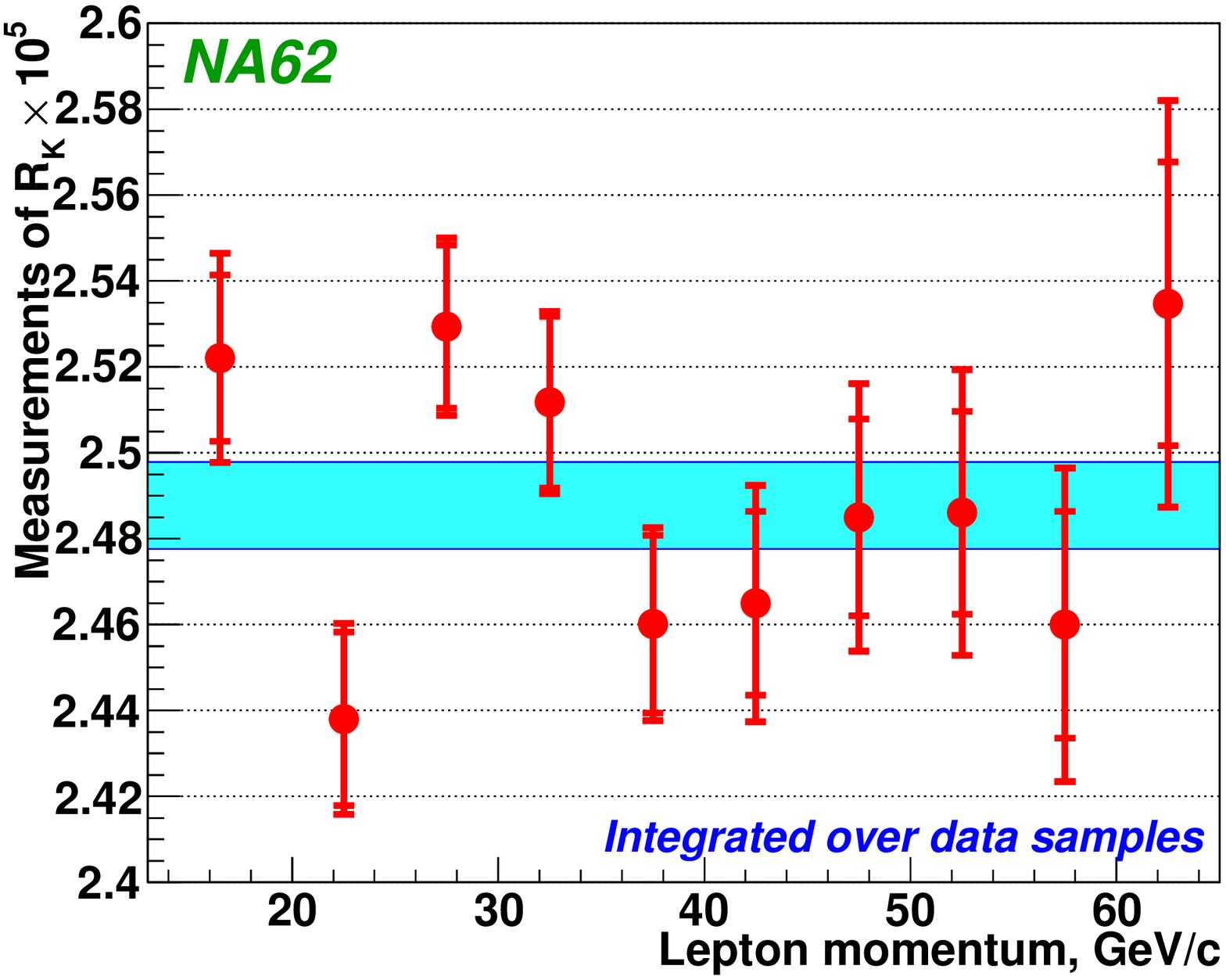}}%
\resizebox{0.4\textwidth}{!}{\includegraphics{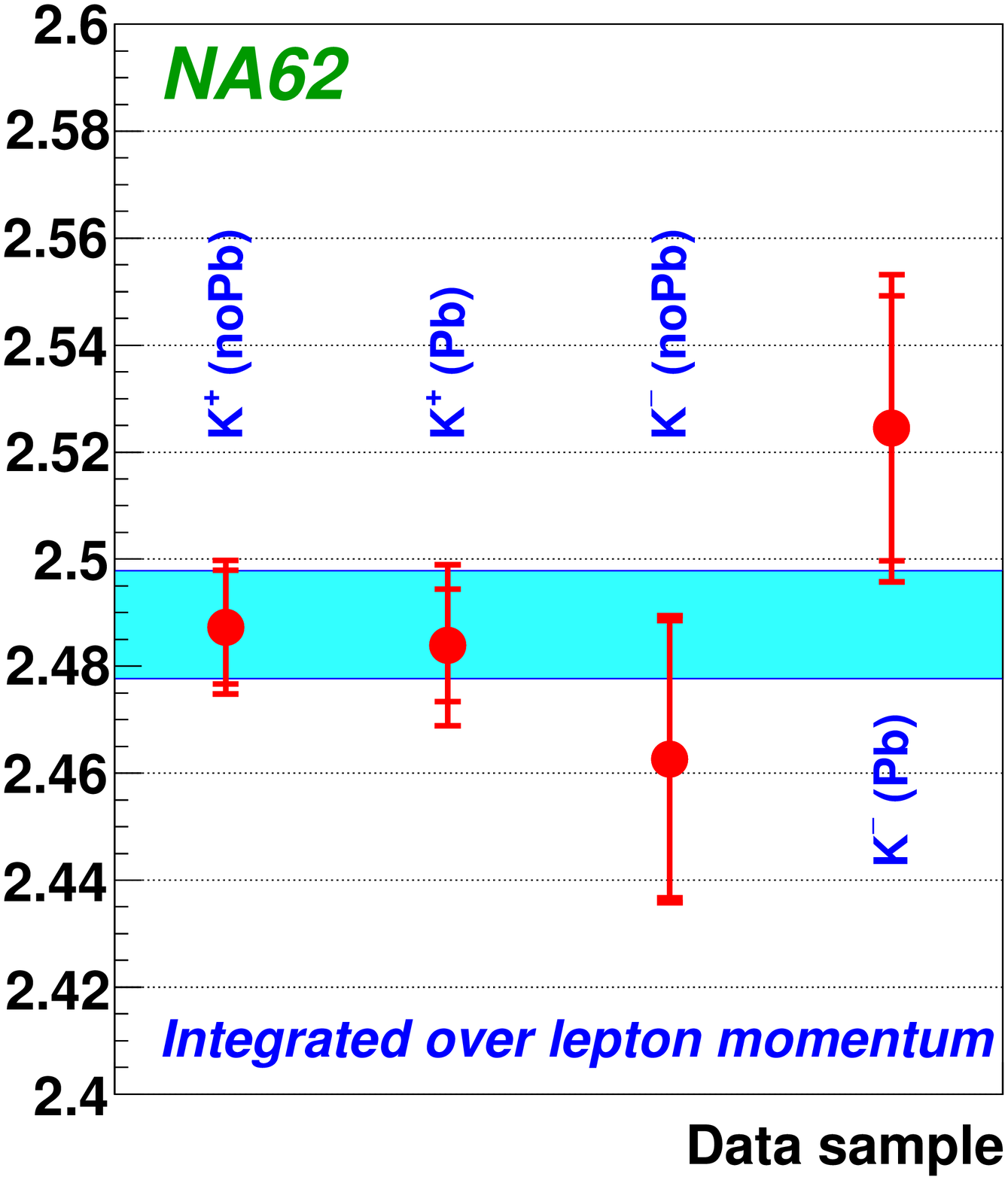}}%
\end{center}
\vspace{-5mm} \caption{Stability of the $R_K$ measurement versus
lepton momentum and for independent data samples. The ``Pb/noPb''
labels indicate samples with present and absent lead wall covering a
part of the LKr calorimeter in order to measure the muon
mis-identification probability. The result of the combined fit over
the 40 data bins is shown by horizontal bands.}\label{fig:rkfit}
\end{figure*}

The numbers of selected $K_{e2}$ and $K_{\mu 2}$ candidates are
145,958 and $4.2817\times 10^7$ (the latter pre-scaled at trigger
level). Backgrounds in the $K_{e2}$ sample integrated over lepton
momentum are summarised in Table~\ref{tab:bkg}: they have been
estimated with Monte Carlo simulations, except for the beam halo
background measured directly with dedicated data samples. The
contributions to the systematic uncertainty of the result include
the uncertainties on the backgrounds, helium purity in the
spectrometer tank (which influences the detection efficiency via
bremsstrahlung and scattering), beam simulation, spectrometer
alignment, particle identification and trigger efficiency. The final
result of the measurement, combined over the 40 independent samples
taking into account correlations between the systematic errors, is
\begin{eqnarray*}
R_K &=& (2.488\pm 0.007_{\mathrm{stat.}}\pm
0.007_{\mathrm{syst.}})\times 10^{-5} =\\
&=&(2.488\pm0.010)\times 10^{-5}.
\end{eqnarray*}
The stability of $R_K$ measurements in lepton momentum bins and for
the separate data samples is shown in Fig.~\ref{fig:rkfit}. The
result is consistent with the Standard Model expectation, and the
achieved precision dominates the world average.

\begin{table}[tb]
\begin{center}
\caption{Summary of backgrounds in the $K_{e2}$ sample.}
\label{tab:bkg}
\begin{tabular}{l|c}
\hline Source & $N_B/N(K_{e2})$\\
\hline $K_{\mu 2}$             & $(5.64\pm0.20)\%$\\
$K_{\mu 2}$ ($\mu\to e$ decay) & $(0.26\pm0.03)\%$\\
$K^\pm\to e^\pm\nu\gamma~(\mathrm{SD}^+)$ & $(2.60\pm0.11)\%$\\
$K^\pm\to\pi^0 e^\pm\nu$           & $(0.18\pm0.09)\%$\\
$K^\pm\to\pi^\pm\pi^0$             & $(0.12\pm0.06)\%$\\
Beam halo                & $(2.11\pm0.09)\%$\\
Decays of opposite sign $K$    & $(0.04\pm0.02)\%$\\
\hline
Total & $(10.95\pm0.27)\%$\\
\hline
\end{tabular}
\end{center}
\end{table}

\section{Search for lepton number conservation violation}

The $K^\pm\to\pi^\mp\mu^\pm\mu^\pm$ decay has been searched for
using a large sample of $K^\pm$ decays with at least three charged
tracks in the final state collected by the NA48/2 experiment in
2003--2004. Three-track vertices with no significant missing
momentum are reconstructed from the magnetic spectrometer
information. Identification of pion and muon candidates is performed
on the basis of energy deposition in the LKr calorimeter and the
muon detector. The muon identification efficiency has been measured
to be above $98\%$ for $p>10$~GeV/$c$.

\begin{figure*}[tb]
\begin{center}
{\resizebox*{0.5\textwidth}{!}{\includegraphics{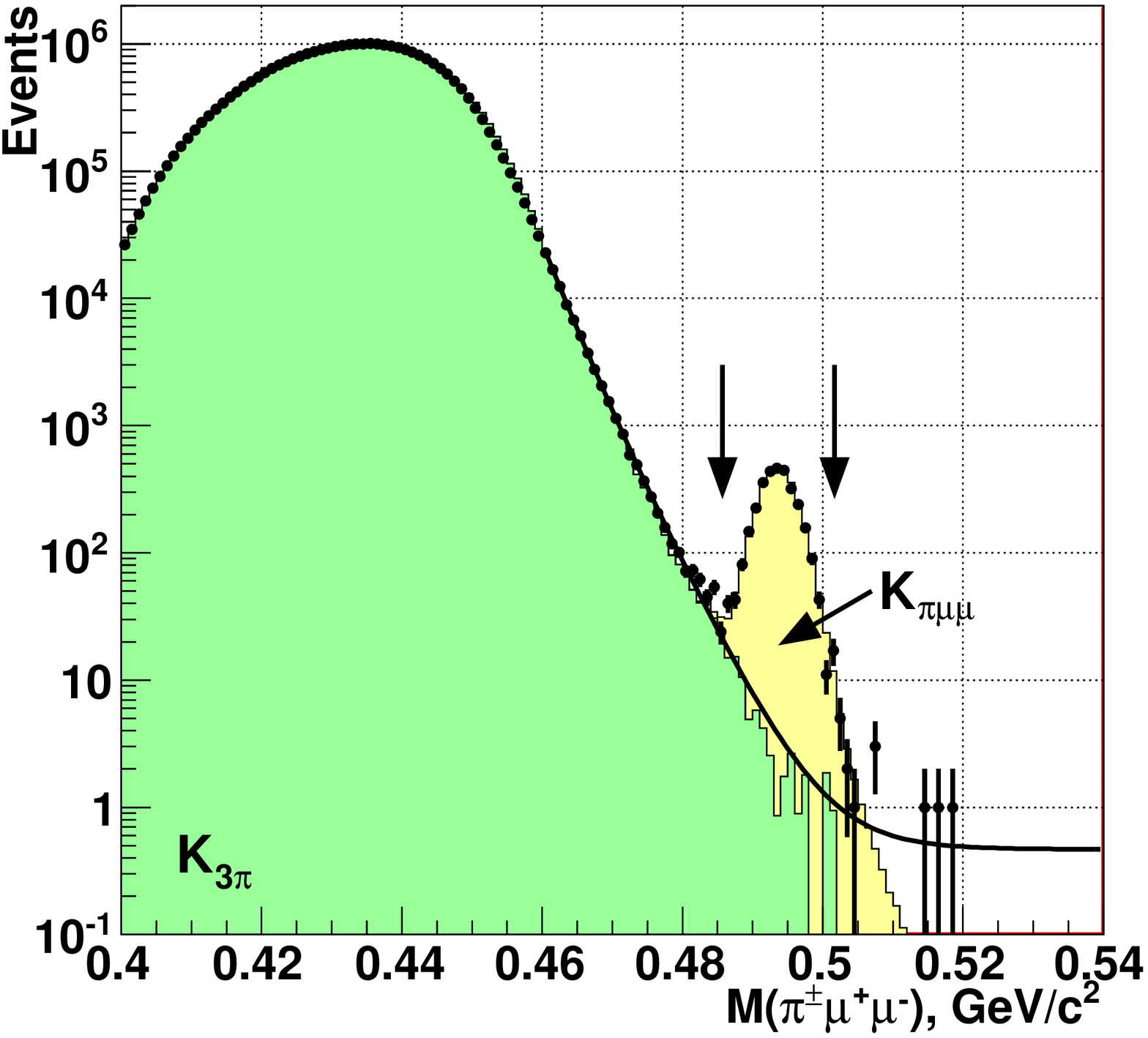}}}%
{\resizebox*{0.5\textwidth}{!}{\includegraphics{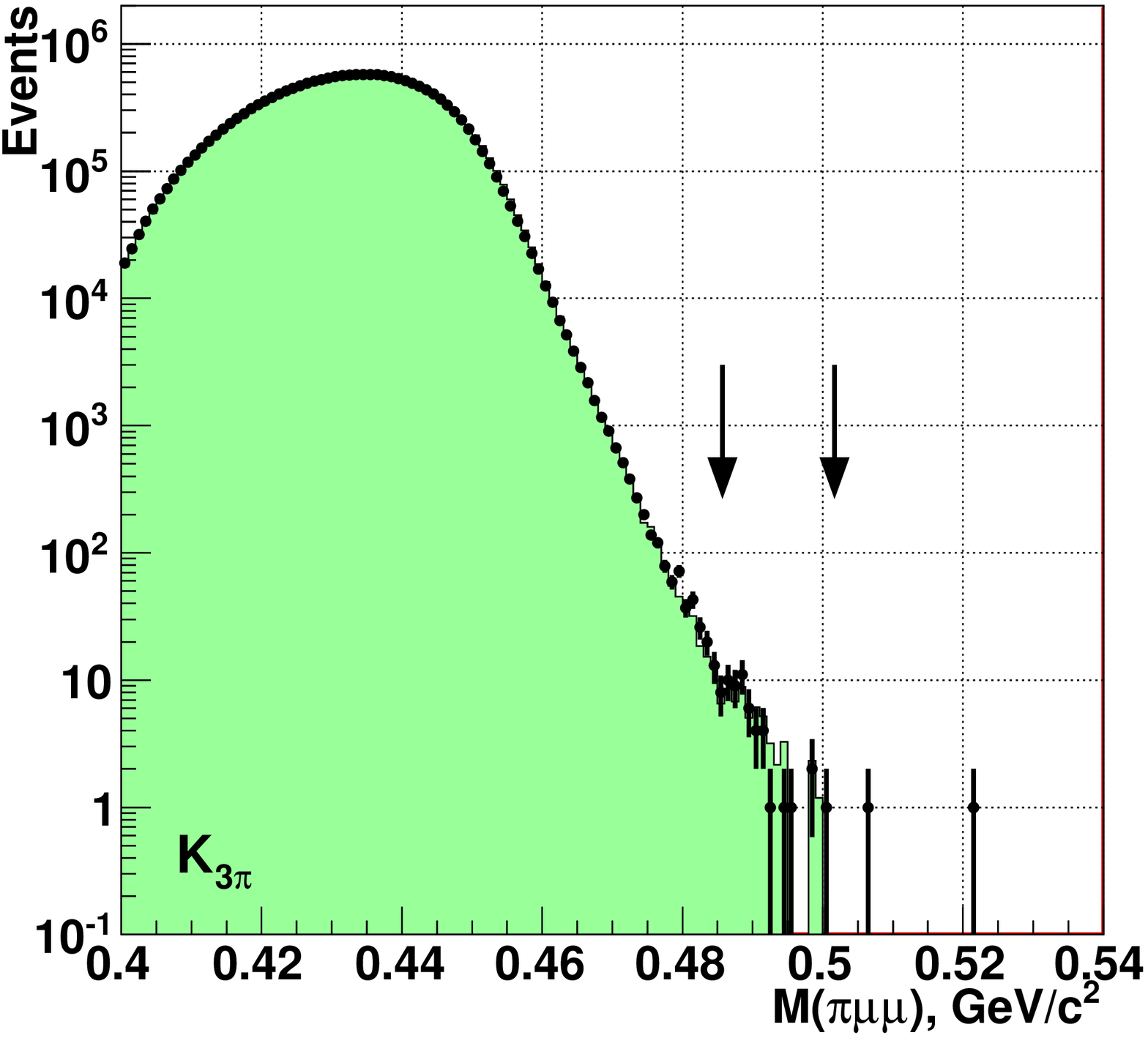}}}%
\end{center}
\vspace{-5mm} \caption{Reconstructed $M_{\pi\mu\mu}$ spectra of
$K^\pm\to\pi^\pm\mu^\pm\mu^\mp$ (left) and lepton flavour violating
$K^\pm\to\pi^\mp\mu^\pm\mu^\pm$ (right) candidates: data (dots),
$K^\pm\to3\pi^\pm$ and $K\to\pi\mu\mu$ MC simulations (filled
areas); fit to background using the empirical parameterization. The
signal regions are indicated with arrows.} \label{fig:lnv}
\end{figure*}

The invariant mass spectrum of the reconstructed
$\pi^\pm\mu^\pm\mu^\mp$ and $\pi^\mp\mu^\pm\mu^\pm$ candidates is
presented in Fig.~\ref{fig:lnv}. The observed
$K^\pm\to\pi^\pm\mu^\pm\mu^\mp$ decay signal has been analyzed
separately~\cite{pimm}. For the lepton number violating signature,
52 events are observed in the signal region
$|M_{\pi\mu\mu}-M_K|<8~{\rm MeV}/c^2$. The background comes from the
$K^\pm\to 3\pi^\pm$ decays with subsequent $\pi^\pm\to\mu^\pm\nu$
decay, is well reproduced by Monte Carlo simulation, and has been
estimated by the simulation to be $(52.6\pm19.8)$ events. The quoted
uncertainty is systematic due to the limited precision of MC
description of the high-mass region, and has been estimated from the
level of agreement of data and simulation in the control mass region
of (465; 485)~MeV/$c^2$. This background estimate has been
cross-checked by fitting the mass spectrum with an empirical
function in the region between 460 and 520~${\rm MeV}/c^2$,
excluding the signal region between 485 and 502~${\rm MeV}/c^2$.

Conservatively assuming the expected background to be
$52.6-19.8=32.8$ events to take into account its uncertainty, the
upper limit for the possible signal is 32.2 events at 90\% CL. The
geometrical acceptance is conservatively assumed to be the smallest
of those averaged over the $K^\pm\to\pi^\pm\mu^\pm\mu^\mp$ and
$K^\pm\to3\pi^\pm$ samples ($A_{\pi\mu\mu}=15.4\%$ and
$A_{3\pi}=22.2\%$). This leads to an upper limit of ${\rm
BR}(K^\pm\to\pi^\mp\mu^\pm\mu^\pm)<1.1\times 10^{-9}$ at 90\% CL,
which improves the best previous limit by almost a factor of 3.
Further details of the analysis are presented in~\cite{pimm}.

\section{The ultra-rare decay $K^+\to\pi^+\nu\bar\nu$}

The principal goal of the NA62 experiment is the measurement of the
branching ratio of the $K^+\to\pi^+\nu\bar\nu$ decay to $\sim 10\%$
precision. The experiment is expected to collect about 100 signal
events in two years of data taking, keeping the systematic
uncertainties and backgrounds low. Assuming a 10\% signal acceptance
and the SM decay rate, the kaon flux should correspond to at least
$10^{13}$ $K^+$ decays in the fiducial volume. In order to achieve a
small systematic uncertainty, a rejection factor for generic kaon
decays of the order of $10^{12}$ is required, and the background
suppression factors need to be measured directly from the data. In
order to achieve the required kaon intensity, signal acceptance and
background suppression, most of the existing NA48/NA62 apparatus
will be replaced with new detectors.

The CERN SPS extraction line used by the NA48 experiment is capable
of delivering beam intensity sufficient for the NA62. Consequently
the new setup will be housed at the CERN North Area High Intensity
Facility where the NA48 was located. The decay in flight technique
will be used; optimisation of the signal acceptance drives the
choice of a 75 GeV/$c$ charged kaon beam with 1\% momentum bite. The
experimental setup is conceptually similar to the one used for NA48:
a $\sim 100$~m long beam line to form the appropriate secondary
beam, a $\sim 80$~m long evacuated decay volume, and a series of
downstream detectors measuring the secondary particles from the
$K^+$ decays in the fiducial decay volume.

The signal signature is one track in the final state matched to one
$K^+$ track in the beam. The integrated rate upstream is about 800
MHz (only 6\% of the beam particles are kaons, the others being
mostly $\pi^+$ and protons). The rate seen by the detector
downstream is about 10 MHz, mainly due to $K^+$ decays. Timing and
spatial information are required to match the upstream and
downstream track.

Backgrounds come from kaon decays with a single reconstructed track
in the final state, including accidentally matched upstream and
downstream tracks. The background suppression profits from the high
kaon beam momentum. A variety of techniques will be employed in
combination in order to reach the required level of background
rejection. They can be schematically divided into kinematic
rejection, precise timing, highly efficient photon and muon veto
systems, and precise particle identification systems to distinguish
$\pi^+$, $K^+$ and positrons. The above requirements drove the
design and the construction of the subdetector systems.

The main NA62 subdetectors are: a differential Cherenkov counter
(CEDAR) on the beam line to identify the $K^+$ in the beam; a
silicon pixel beam tracker; guard-ring counters surrounding the beam
tracker to veto catastrophic interactions of particles; a downstream
spectrometer composed of straw chambers operating in vacuum; a RICH
detector to distinguish pions and muons; a scintillator hodoscope; a
muon veto detector. The photon veto detectors will include a series
of annular lead glass calorimeters surrounding the decay and
detector volume, the NA48 LKr calorimeter, and two small angle
calorimeters to keep the hermetic coverage for photons emitted at
close to zero angle to the beam.

The design of the experimental apparatus and the R\&D of the new
subdetectors are complete. The experiment is under construction, and
the first technical run is scheduled for October--December 2012.

\section{Conclusions}

The NA48/NA62 experiments at CERN recently accomplished a series of
precision measurements of rare $K^\pm$ decays. The $R_K$ phase of
the NA62 experiment provided the most precise measurement of the
lepton flavour parameter $R_K=(2.488\pm0.010)\times 10^{-5}$. This
result is consistent with the SM expectation, and constrains
multi-Higgs and fourth generation new physics scenarios. NA48/2 has
improved the upper limit on the branching ratio of the lepton number
violating decay $K^\pm\to\pi^\mp\mu^\pm\mu^\pm$, which is now
$1.1\times 10^{-9}$ at 90\% CL. The ultra-rare
$K^+\to\pi^+\nu\bar\nu$ decay represents a unique environment to
search for new physics. The NA62 experiment, aiming to collect
${\cal O}(100)$ events of this decay, is being constructed and is
preparing for a technical run in 2012.



\nocite{*}
\bibliographystyle{elsarticle-num}
\bibliography{martin}



\end{document}